\documentclass[12pt]{article}
\usepackage{amsfonts}

\newcommand{\pp}[2]{\mbox{pp.~#1\,--\,#2}}

\newcommand{\oosrt}{\squash{\frac{1}{\sqrt{2}}}}
\newcommand{\ket}[1]{\mbox{$|#1\rangle$}}
\newcommand{\squash}[1]{\raisebox{0.04ex}[0pt][0pt]{\small$\textstyle #1$}}
\newcommand{\pbfrac}[2]{\mbox{$\mbox{}^{#1}\!/_{#2}$}}
\newcommand{\pbhalf}{\mbox{$\textstyle \pbfrac{1}{2}$}}
\newcommand{\eps}{\varepsilon}
\newcommand{\oneovereps}{{\frac{1}{\raisebox{0.35ex}{\scriptsize $\eps$}}}}

\newcommand{\dee}{\mbox{\rm d}}
\newcommand{\half}{\squash{\frac{1}{2}}}

\title{\textbf{Quantum Communication Complexity} \\
\large\bf (a survey)}

\author{\large Gilles Brassard\,%
\thanks{\,Supported in part by Canada's {\sc Nserc} and Qu\'ebec's {\sc Fcar}.}
\\ {\normalsize\it D\'epartement~IRO, Universit\'e de Montr\'eal}\\[-1ex]
{\normalsize\it C.P.~6128, succursale centre-ville}\\[-1ex]
{\normalsize\it Montr\'eal (Qu\'ebec), Canada H3C~3J7}\\ 
{\normalsize\texttt{brassard}\textbf{\char"40}\texttt{iro.umontreal.ca}}}
\date{{\normalsize 15 December 2000}}

\begin{document}
\maketitle
\thispagestyle{empty}

\begin{abstract}

Can quantum communication be more efficient than its classical counterpart?
Holevo's theorem rules out the possibility of communicating more than $n$
bits of classical information by the transmission of $n$ quantum bits---%
unless the two parties are entangled, in which case twice as many classical
bits can be communicated but no more.  In~apparent contradiction,
there are distributed computational tasks for which quantum communication
cannot be simulated efficiently by classical means.  In~extreme cases,
the effect of transmitting quantum bits cannot be achieved classically
short of transmitting an exponentially larger number of bits.

In a similar vein, can entanglement be used to save on classical communication?
It~is well known that entanglement on its own is useless for the transmission
of information.  Yet, there are distributed tasks that cannot be
accomplished at all in a classical world when communication is not allowed,
but that become possible if the non-communicating parties share prior
entanglement.  This leads to the question of how expensive it is,
in terms of classical communication, to provide an exact simulation of the
spooky power of entanglement.

\end{abstract}

\section{Introduction}

It is well known that the use of quantum information allows for
tasks that would be provably impossible in a classical world, such as
the transmission of unconditionally confidential information between
parties that share only a short secret key~\cite{BB84,sciam}.
Apart from this \emph{quantum cryptography}, are there advantages to be gained
in setting up an infrastructure that would facilitate the transmission
of quantum information?  In~particular, are there advantages to be
gained in terms of communication \emph{efficiency}?  A~different but related
question concerns quantum entanglement: can entangled parties make better use
of a classical communication channel than their non-entangled counterparts?
Better still: can entangled parties benefit from their entanglement
if they are not allowed any form of direct communication?

There are good reasons to believe at first that the answer to all the
above questions is negative.  In~particular, Holevo's theorem~\cite{holevo}
states that no more than $n$ bits of expected classical information can be
communicated between unentangled parties by the transmission of $n$
quantum bits---henceforth called \emph{qubits}---regardless of the coding
scheme that could be used. If~the communicating parties share prior
entanglement, twice as much classical information can be
transmitted~\cite{densecoding}, but no more. This applies even if the
communication is not restricted to be \mbox{one-way}~\cite{CvDNT}.
It~is thus reasonable to expect that no significant savings in communication
can be achieved by the transmission of qubits, and possibly no savings
at all if the communicating parties do not share prior entanglement.
As~for the last question, it is well known that entanglement alone cannot
be used to signal information---otherwise faster-than-light communication
would be possible and causality would be violated---and thus it would seem
that entanglement is useless if it is not supplemented by direct forms of
communication. Here~we survey striking results to the effect that all the
intuition in this paragraph is wrong.

After a review of classical communication complexity in Section~\ref{class},
we consider in Section~\ref{alayao} the situation in which quantum
communication is allowed.  In~Section~\ref{alaCB}, we revert to
classical communication but allow unlimited prior entanglement between the
communicating parties.  Section~\ref{spooky} investigates in more detail the
power of prior entanglement when no direct communication is allowed to take
place, which we call spooky communication complexity. In~Section~\ref{simul},
we determine how expensive it is to simulate the effect of entanglement in a
purely classical world.
Finally, we conclude with open problems in Section~\ref{concl}.
Although we do not cover the important topic of lower
bounds for quantum communication complexity, we encourage the reader to
consult~\mbox{\cite{kremer,CvDNT}} for early results and~\cite{BdW}
for powerful new techniques.

\section{Classical Communication Complexity}\label{class}

Let~\mbox{Alice} and Bob be two distant parties who wish to collaborate
on a common task that depends on distributed inputs.
More precisely, let $X$, $Y$ and $Z$ be sets
and consider a function \mbox{$f:X \times Y \rightarrow Z$}.
Assume Alice and Bob are given some
\mbox{$x \in X$} and \mbox{$y \in Y$}, respectively, and their goal is to
compute \mbox{$z=f(x,y)$}.
Sometimes, we add a \emph{promise} \mbox{$P(x,y)$}
for some Boolean function~$P$, in which case Alice and Bob are required
to compute the correct answer \mbox{$f(x,y)$} only when~\mbox{$P(x,y)$} holds.
Whether or not there is a promise, the obvious recipe is for Alice to
communicate $x$ to Bob, which allows him to compute~$z$.
Once obtained, Bob can then communicate $z$ back to Alice if both
parties need to know the answer.
If~we are concerned with the amount of communication required to
achieve this task---paying no attention to the computing effort involved
in the process---could there be more efficient solutions for some
functions~$f$\,? For~\emph{all} functions?

The answer is obviously positive for the first question.  For~example,
if \mbox{$X=Y=\{0,1\}^n$} for some integer~$n$, \mbox{$Z=\{0,1\}$}
and \mbox{$f(x,y)$} is defined to be 0 if and only if $x$ and $y$ have the
same Hamming weight (the same number of~1s), it suffices for Alice to
communicate the Hamming weight of $x$ to Bob for him to verify the condition.
Thus, about $\lg n$ bits\,\footnote{\,The symbol ``$\lg$''~is used to denote the
base-two logarithm.} of communication are sufficient for this task,
which is much more economical than if Alice had transmitted all $n$ bits
of her input to Bob.  The~\mbox{answer} to the second question, however, is
negative: There are functions for which the obvious solution \emph{is} optimal. 
For~instance, $n$ bits of communication are necessary and sufficient in the
worst case for Bob to decide whether or not Alice's input is the same as~his.
This unsurprising statement is not easy to prove, but a host of
techniques have been developed to handle that kind of questions.
See~\cite{KN97} for a survey.

A more interesting scenario takes place when we do not insist on the correct
\mbox{answer} to be obtained with certainty.  If~we accept a small \mbox{error}
probability, we can do significantly better on the above-mentioned
equality-testing problem.
Let~\mbox{$\eps>0$} be the tolerated error probability and let
$p$ be the smallest prime number larger than~$n/\eps$.
Let $\mathbb F$ be the finite field with $p$ elements.
Upon receiving their inputs $x$ and~$y$, \mbox{Alice} forms polynomial
\mbox{$P(z)=x_1+x_2 z+x_3 z^2+\cdots+x_n z^{n-1}$} over $\mathbb F$
and Bob forms \mbox{$Q(z)=y_1+y_2 z+y_3 z^2+\cdots+y_n z^{n-1}$}.
Then, Alice chooses a random element \mbox{$w \in \mathbb F$}.
She~computes $v=P(w)$ and transmits both $w$ and $v$ to~Bob, who
computes $Q(w)$ and compares the answer with~$v$.
If~\mbox{$Q(w) \neq v$}, it has been established that \mbox{$x \neq y$}.
\mbox{Otherwise}, Bob can claim with confidence that \mbox{$x=y$} because
two distinct polynomials of degree smaller than~$n$ cannot agree on more
than~$n$ distinct points and therefore the proportion of points in~$\mathbb F$
on which $P(z)$ and $Q(z)$ agree must be less than
\mbox{$n/\#{\mathbb F} = n/p < \eps$}.
Note that \mbox{$p \leq 2 n/\eps$} since there is always a prime number
between any number and its double, and hence each of $w$ and $v$
can be written with no more than \mbox{$2+\lg n + \lg \oneovereps$} bits.
The~communication complexity of this protocol is therefore at most twice
this many bits, which is much less than $n$ for any fixed error probability
$\eps$ when $n$ is large enough.

In the classical model of communication complexity, it is often allowed for
Alice and Bob to \emph{share random variables} even though one may argue that
this does not make much sense from a mathematical point of view.  In~this
scenario, we assume that Alice draws a random bit string (or integer) according
to some specific distribution---or~sometimes even a random real number---and
she tells Bob the outcome of this draw in an \emph{initialization phase}.
This communication is not accounted for in the complexity of the protocol
because it takes place \emph{before} Alice and Bob are given their respective
inputs.  When correctness of the protocol is analysed for a given input,
probabilities are taken over the possible choices of that random variable, as
if it had been chosen \emph{after} the inputs were determined\,%
\footnote{\,Note that drawing shared
random bits once Alice and Bob are separated makes perfect sense in a quantum
world if they share entanglement: they simply have to measure corresponding
qubits from
\vspace{1ex}\mbox{$\ket{\Phi^{+}}=\oosrt\ket{00}+\oosrt\ket{11}$}
pairs in the computational basis.}.
In~this model, the equality-testing problem can be solved with error probability
$\eps$ with only \mbox{$m=\lceil \, \lg 1/\eps \rceil$} bits of
communication, regardless of the value of~$n$.
In~the initialization phase, Alice and Bob share $m$ random bit strings
\mbox{$a_1$, $a_2$, \ldots, $a_m$}\,, each of length~$n$.  Once they receive
their inputs $x$ and~$y$, Alice computes \mbox{$b_i=x \cdot a_i$} for each~$i$,
where $x \cdot a$ is the \emph{inner product}\,%
\footnote{\,To compute the inner product between two bit strings of equal
length, line them up one under the other and count the number of positions
in which they both have a~1.  If~this number is even, the inner product
is~0; otherwise it is~1.  Mathematically, it is the exclusive~or (sum modulo~2)
of the bitwise \textsc{and} of the two strings.}
between bit strings $x$ and~$a$.
Alice transmits \mbox{$b_1$, $b_2$, \ldots, $b_m$} to Bob,
who verifies whether or not \mbox{$b_i=y \cdot a_i$} for each~$i$.
If~not, it has been established that \mbox{$x \neq y$}.
Otherwise, Bob can claim with confidence that \mbox{$x=y$} because
the probability of error of this strategy is \mbox{$2^{-m} \leq \eps$} since it
is~$\pbhalf$ independently for each~$i$.

\newpage

\section{Quantum Communication Complexity}\label{alayao}

The topic of classical communication complexity was introduced and
first studied by Andrew Yao in 1979~\cite{yaoc}.
Almost 15 years elapsed before the same pioneer thought of asking
how the situation might change if Alice and Bob were allowed to
exchange quantum rather than classical bits~\cite{yaoq}.
It~seems at first that no savings in communication are to be expected at all
because of Holevo's theorem~\cite{holevo}, which states that no more than $n$
bits of expected classical information can be communicated between unentangled
parties by the transmission of $n$ qubits.
(It~was implicit in Yao's original model that Alice and Bob were not
allowed to share prior entanglement in the initialization phase.)

The first hint that quantum communication could be more efficient than
classical communication was given in August~1997 by Richard Cleve, Wim van Dam,
Michael Nielsen and Alain Tapp~\cite{CvDNT} in a probabilistic setting\,%
\footnote{\,For~the sake of historical completeness, it is easy to modify
a protocol given three months earlier by Buhrman, Cleve and
van~Dam~\cite{BCvD} (original \texttt{quant-ph} version)
to achieve a similar goal, and indeed this is done in the final version
of that paper, which is to appear in \textit{SIAM Journal on Computing}.}.
Alice and Bob are given two-bits vectors $x_1 x_2$ and $y_1 y_2$,
respectively.  They must both decide if \mbox{$x_1 y_1 + x_2 y_2$}
is even or odd, and they are restricted to two bits of
communication.  Shared random variables are allowed.
It~is proven in~\cite{CvDNT} that no classical protocol for this task can
give the correct answer with a probability better than~\pbfrac{7}{9}.
Yet, if two quantum bits of communication are allowed, instead of two
classical bits, the correct answer can be obtained with an improved
probability of~\pbfrac{4}{5}.

A more convincing case for the superiority of quantum communication
came the following year when Harry Buhrman, Richard
Cleve and Avi Wigderson~\cite{BCW} proved that quantum communication can
be \emph{exponentially} better than classical communication in the error-free
model, provided the inputs respect a  given promise, and almost quadratically
better in the bounded-error promise-free model. Subsequently, Ran Raz proved
that an exponential advantage exists to quantum communication even in the
bounded-error promise-problem model~\cite{raz}.

The first exponential separation~\cite{BCW} was inspired
by the \mbox{famous} Deutsch--Jozsa \mbox{problem}~\cite{DJ}, which was used
in 1992 to show for the first time that quantum computers could be
exponentially faster than classical computers in an oracle-based~\cite{oracle}
error-free setting. More specifically, Buhrman, Cleve and Wigderson defined
the \mbox{following} \mbox{scenario}, where $\Delta(x,y)$ denotes
the \emph{Hamming distance} between bit strings $x$ and~$y$, which is the
number of bit \mbox{positions} on which they differ.
Let~$k$ be an integer, \mbox{$n=2^k$}, \mbox{$X=Y=\{0,1\}^n$}
and \mbox{$Z=\{0,1\}$}.
Function \mbox{$f:X \times Y \rightarrow Z$} is the equality function:
\mbox{$f(x,y)=1$} if and only if \mbox{$x=y$}.  We~have seen already that
this communication complexity problem \mbox{requires} $n$ bits of classical
communication if errors are not tolerated.  Now, we introduce the
promise $P(x,y)$, which is defined to be \textsf{true} if and only if
\mbox{$\Delta(x,y) \in \{0,n/2\}$}.
In~other words, $P(x,y)$ holds if and only if either \mbox{$x=y$}
or $x$ and $y$ differ on exactly half their positions.
It~is proven in~\cite{BCW} that the error-free equality-testing problem
requires at least $cn$ classical bits of communication, for some real
\mbox{positive} constant~$c$ and all sufficiently large~$n$, even when the
correct answer is required only when the promise holds.  (The~hard part of that
proof is taken from~\cite{FR}.)  Even though quantum communication cannot be
significantly more efficient than classical communication for the straight
equality-testing problem~\cite{BdW},
it is shown in~\cite{BCW} that it can be solved with certainty using
as few as $k$ quantum bits of communication whenever the
promise holds, which is exponentially better than the
$cn$ bits that would be required in a classical scenario.
A~quantum protocol for this problem is easily derived from the
first example of ``spooky communication'' given in Section~\ref{spooky}.

This exponential ``superiority'' of quantum over classical communication
is not \mbox{entirely} convincing because it vanishes as soon as we tolerate
an arbitrarily small probability of error.  Indeed, we have seen that the
equality-testing problem can be solved with a \emph{constant} number of bits
of classical communication, for any fixed error probability, when shared
random variables are allowed\,\footnote{\,Even if we do not allow shared
random variables, the problem can be solved with \mbox{$2k+c$} bits
of classical communication, where \mbox{$k=\lg n$} and $c$ is a constant that
depends only on the error probability.}.  The other
problem featured in~\cite{BCW} is more interesting, even though the quantum
superiority is merely \mbox{almost} quadratic, because it applies in the
more realistic bounded-error model.  Consider the following scenario. 
Alice and Bob are very busy and they would like to find a time when they are
simultaneously free for lunch. They each have an engagement calendar, which we
think of as an \mbox{$n$--bit} string $x$ (resp~$y$), where \mbox{$x_i=1$}
(resp.~\mbox{$y_i=1$}) means that Alice (resp.~Bob) is free for lunch on
day~$i$. Mathematically, they want to find an index~$i$ such that
\mbox{$x_i=y_i=1$} or establish that such an index does not exist.
Balasubramanian Kalyanasundaram and Georg Schnitger~\cite{KS} proved in 1987
that this task requires at least $cn$ classical bits of expected communication
in the worst case, for some real positive constant~$c$ and all sufficiently
large~$n$, even when the answer is only required to be correct with
probability at least~\pbfrac{2}{3}. Intuitively, this means that lunch cannot be
scheduled short of exchanging a constant fraction of the appointment calendar.
In~sharp contrast, it is shown in~\cite{BCW} that this problem can
be solved with the exchange of at most \mbox{$d \sqrt{n} \lg{n}$} quantum
bits for some constant $d$ and all sufficiently large~$n$.
This is accomplished by implementing a distributed version of Grover's
quantum search algorithm~\cite{grover} in which we search for a~$1$ in the
bitwise \textsc{and} of $x$ and~$y$.  A~\mbox{$\lceil\,\lg n\rceil$--qubit}
quantum register is shuttled back and forth between Alice and Bob for each
of the approximately $\sqrt{n}$ iterations of Grover's algorithm.

\section{Substituting Entanglement for Communication}\label{alaCB}

A slightly different model was introduced by Richard Cleve and Harry
Buhrman~\cite{CB}.
\mbox{Assume} Alice and Bob are restricted to communicating classical
information. Are~there tasks for which they could save on the required amount of
communication if they share prior entanglement?
Again, it is tempting to think that this is not possible because
entanglement cannot be used to increase the capacity of a classical channel:
Alice cannot communicate more than $n$ expected bits of
classical information to Bob if less than $n$ bits are actually transmitted
between them---even if they are allowed two-way communication and unlimited
use of entanglement.  And again, this intuition is wrong.

In their original paper~\cite{CB}, Cleve and Buhrman were able to show that
entanglement can be used to save one bit of classical communication, but only
in a three-party scenario.  Still, this was the very first example of a
distributed task that could be solved more efficiently (in~terms of
communication) in our quantum world than would be possible in a sad
classical world, because it predates~\cite{CvDNT} by four months.
Subsequently, Buhrman, Cleve and van~Dam~\cite{BCvD} found a
\emph{two-party} distributed problem that can be solved with a
probability of success exceeding $85\%$ if prior shared entanglement
is available, whereas the probability of success in a classical world
could not exceed $75\%$ with the same amount of communication, even if
shared random variables are allowed.

The first problem for which communication complexity could be reduced by more
than a constant additive amount was also discovered in this shared-entanglement
\mbox{scenario}, rather than in Yao's original qubit-transmission scenario
described in the previous section: Harry Buhrman, Wim van Dam, Peter H{\o}yer
and Alain Tapp~\cite{BvDHT} gave a \mbox{$k$--party} distributed task that
requires roughly \mbox{$k \lg k$} bits of communication in a classical world,
yet it can be carried out with exactly $k$ bits of classical communication if
the parties are allowed to share prior entanglement.

The exponential and almost-quadratic improvements mentioned
in the \mbox{previous} section~\cite{BCW,raz} apply just as well in the
shared-entanglement scenario.
This is obvious since the effect of any protocol that requires the
communication of $\ell$ quantum bits can be achieved
by the transmission of~$2\ell$ classical bits---through
\emph{quantum teleportation}~\cite{teleport}---provided $\ell$ bits of shared
entanglement are available.

There is yet another natural communication complexity scenario, in which
the \mbox{parties} are allowed to share prior entanglement \emph{and} to
communicate quantum bits.  For~the sake of brevity, we shall not
elaborate on this approach here.

\section{Spooky Communication Complexity}\label{spooky}

An even more intriguing question is to determine if entanglement can
be used \emph{instead~of} communication.  Are there tasks that would be
impossible to achieve in a classical world if Alice and Bob were not
allowed to communicate, yet those tasks can be performed without \emph{any}
form of communication provided the participants share prior entanglement?
In~the words of Alain Tapp, this would provide a form of
\emph{pseudo telepathy} because it would give the \emph{illusion} of
communication between Alice and Bob when in fact no such communication takes
place. And indeed there would be no communication because entanglement
cannot be used to signal information: nothing Alice can do locally on
her quantum system can cause a measurable change in Bob's,
no matter how they are entangled.

A moment's thought suffices to realize that pseudo telepathy is possible if we
are content with probabilistic tasks.
Define the \emph{EPR~task} as follows. Once separated, Alice and Bob are
given each a real number $x$ and~$y$, respectively, between 0 and~$\pi$. 
They are to produce each a single bit: $a$~for Alice and $b$ for Bob. 
Alice's output must be equally likely to be 0 or~1, and so must Bob's output,
but the required correlation is that \mbox{$a=b$} with probability
\mbox{$\cos^2(x-y)$}.  It~is precisely the essence of Bell's
theorem~\cite{bell} that such correlations cannot be established in a
classical world if communication between Alice and Bob is not allowed,
even if the inputs are restricted to binary choices \mbox{$x \in \{0,\pi/6\}$}
and \mbox{$y \in \{0,5\pi/6\}$}.
Yet, it is easy for participants who share a
\mbox{$\ket{\Phi^{+}}=\oosrt\ket{00}+\oosrt\ket{11}$} state
to realize this task.  If~we think of this state as a pair of entangled
polarized photons, it suffices for Alice and Bob to measure their photons at
polarization angles~$x$ and~$y$, respectively, and the outcomes of the
measurements provide the required outputs $a$ and~$b$. 

But is pseudo telepathy possible when there is a deterministic criterion
to decide if the goal has been achieved and when errors are not tolerated?
This brings us to our last form of communication complexity, which we call
\emph{spooky communication \mbox{complexity}}.
As~usual, let $X$, $Y$ and $Z$ be sets
and consider a function \mbox{$f:X \times Y \rightarrow Z$}
such that it is not possible to compute the value of $f(x,y)$ with
certainty from knowledge of either $x$ or~$y$ alone.  It~follows that
if Alice and Bob are given \mbox{$x \in X$} and \mbox{$y \in Y$},
respectively, they cannot compute $f(x,y)$ without communication.
Can such a function $f$ exist so that Alice and Bob---or~at least
one of them---could compute $f(x,y)$ nevertheless \mbox{provided} the
participants share prior entanglement?  Of~course not, since this
would allow for faster-than-light communication!  Thus, we have
to define spooky communication \mbox{complexity} in a more subtle way,
through a \emph{relation} rather than a function.

Let $X$, $Y$, $A$ and $B$ be sets
and consider a relation \mbox{$R \subseteq X \times Y \times A \times B$}.
In~an \emph{initialization phase}, Alice and Bob are allowed to discuss
strategy and share random variables.  They are also allowed to share
entanglement.  After Alice and Bob are physically separated, they are given
\mbox{$x \in X$} and \mbox{$y \in Y$}, respectively.
Without being allowed any forms of communication, their
goal is to produce \mbox{$a \in A$} and \mbox{$b \in B$}, respectively,
such that \mbox{$(x,y,a,b) \in R$}.  We~say that \emph{spooky communication},
or~\emph{pseudo telepathy}, takes place if this task could not be fulfilled
with certainty in a classical world, whereas it can provided Alice
and Bob share prior entanglement.  The~amount of spooky communication
\emph{complexity} is measured in the number of bits of entanglement
that are required to succeed in the worst case.
The~\emph{spooky advantage} is defined as the function that relates the
spooky complexity to the number of classical bits of communication that
would be needed in the worst case to succeed in the classical setting.

The first example of spooky communication was provided by Gilles Brassard,
Richard Cleve and Alain Tapp~\cite{BCT} as yet another variation on the
Deutsch--Jozsa problem~\cite{DJ}.  Let~$k$ be an integer, \mbox{$n=2^k$},
\mbox{$X=Y=\{0,1\}^n$} and \mbox{$A=B=\{0,1\}^k$}.  The~\emph{Deutsch--Jozsa
\mbox{relation}} $R$ is defined as follows, where $\Delta(x,y)$ denotes again
the Hamming distance between $x$ and~$y$.
\[ (x,y,a,b) \in R ~~\Longleftrightarrow~~ \left\{
\begin{array}{l}
x=y \mbox{ and } a=b, \mbox{ or} \\[1ex]
\Delta(x,y)=n/2 \mbox{ and } a \neq b, \mbox{ or} \\[1ex]
\Delta(x,y) \not\in \{0,n/2\}
\end{array}  \right.
\]
In other words, Alice and Bob are promised that either their inputs
are the same, or that they differ on exactly half the bits.  They must produce
identical outputs if and only if their inputs are the same.  But if the promise
is not fulfilled, there are no conditions on what Alice and Bob produce.
The challenge comes from the fact that the outputs $a$ and $b$ must be
exponentially shorter than the inputs $x$ and~$y$.

It is not immediate that the Deutsch--Jozsa relation requires
communication to be established in the classical setting.  After all,
it \emph{can} be established when \mbox{$n=2$}
(easily) and~\mbox{$n=4$} (think about it!).  But~it is proven in~\cite{BCT}
that there exists a positive constant $c$ such that the Deutsch--Jozsa relation
cannot be \mbox{established} classically with fewer than $cn$ bits of
communication provided $n$ is sufficiently large, based on the similar lower
bound from~\cite{BCW} that we had mentioned in Section~\ref{alayao}.
On~the other hand, we show below that the Deutsch--Jozsa relation \emph{can}
be established in the spooky setting with as few as $k$ bits of entanglement. 
It~follows that the spooky advantage of this problem is exponential because $n$
is exponential in~$k$.

To establish the Deutsch--Jozsa relation, Alice creates a \mbox{$2k$-qubit}
register in state
\[ \sum_{z \in \{0,1\}^k} \, 2^{-k/2} \, \ket{z,z} \, , \]
which is the same as $k$ pairs in state \ket{\Phi^{+}} up to ordering of
the qubits.
She keeps the first $k$ qubits of that register and gives the other
$k$ qubits to Bob.  After Alice and Bob are separated, they receive
their respective inputs $x$ and~$y$.  To~each integer~$i$,
\mbox{$1 \le i \le n$}, associate the bit string \mbox{$z_i \in \{0,1\}^k$}
that represents number \mbox{$i-1$} in binary.
Now, Alice applies to her register the unitary transformation that maps
\ket{z_i} to \mbox{$(-1)^{x_i} \, \ket{z_i}$} for each~$i$, and Bob does the
same to his register, but with $y_i$ instead of~$x_i$.
This produces joint state
\looseness=-1
\[ \sum_{i=1}^{n} \, 2^{-k/2} \, (-1)^{x_i} (-1)^{y_i} \ket{z_i,z_i}
~=~  \sum_{i=1}^{n} \, 2^{-k/2} \, (-1)^{x_i \oplus y_i} \ket{z_i,z_i} \, .
\]
Next, Alice applies the Walsh--Hadamard transform,
which sends \ket{0} to \mbox{$\oosrt \ket{0} + \oosrt \ket{1}$}
and \ket{1} to \mbox{$\oosrt \ket{0} - \oosrt \ket{1}$},
to each of the $k$ qubits of her register, and Bob
does the same on his register.
Finally, Alice and Bob measure their registers in the
computational basis.  The~resulting classical strings, $a$ and~$b$,
are their final output.  It~is proven in~\cite{BCT} that this
process accomplishes the required job.

\section{Classical Simulation of Entanglement}\label{simul}

Once we have established that pseudo telepathy is possible, the next
natural question is to determine how much classical communication is
necessary and sufficient to simulate the effect of $k$ bits of
entanglement.  It~follows from the previous section (and Bell's
theorem~\cite{bell} for small values of~$k$) that at least
$c2^k$ bits are required, for some constant \mbox{$c>0$}
and all \mbox{$k \ge 1$}.  But are these many bits of classical communication
\emph{sufficient} to simulate \emph{everything} that can be accomplished
with $k$ bits of entanglement?

In fact, can the effect of entanglement be simulated \emph{at all} with
a finite amount of classical communication?  As~the simplest possible example,
can the EPR task, as defined in Section~\ref{spooky}, be simulated by
classical communication?  In~particular, we must have \mbox{$a=b$}
if \mbox{$x=y$} and \mbox{$a \neq b$} if \mbox{$|x-y|=\pi/2$}.
Surely, it is not possible for Alice to communicate
her input $x$ to Bob, for this would require an infinite amount of
communication.
Yet, it is shown in~\cite{BCT} that \emph{four} bits of classical communication
are sufficient in the worst case for an exact simulation of the EPR task,
provided Alice and Bob are allowed to share a continuous real random variable
in the initialization phase---admittedly an unreasonable proposition. 
The essence of the idea is best explained if we further restrict the inputs $x$
and $y$ to be between $0$ and~$1$, rather than between $0$ and~$\pi$.  In~this
case, \emph{a~single bit} of classical communication suffices to simulate the
EPR task exactly. This restriction is somewhat bizarre since it translates
to requiring the angles to be between $0$ and approximately $57.3$ degrees.
Without this restriction, a rather painful piecewise construction has to be
implemented, as explained in~\cite{BCT}, and we need four bits of classical
communication to take care of the various possible cases.

In the initialization phase, Alice and Bob share a boolean variable~$c$,
which is equally likely to be 0 or~1, and a continuous real variable $r$
chosen uniformly in the interval~\mbox{$(0,1)$}.  After they are separated,
they receive their angles $x$ and $y$, respectively.
Alice outputs \mbox{$a=c$}, a random bit as required.
Then, she tells Bob if~\mbox{$a<x$} with a single bit of classical
communication.  This allows Bob to determine whether or not $r$ lies
in between $x$ and~$y$.  In~case it does not, Bob outputs the same
bit \mbox{$b=c$} as Alice.  Note in particular that if $x=y$, then we get
$a=b$ with certainty, as required.
On~the other hand, if $r$ does lie between $x$ and $y$, then
Bob outputs \mbox{$b=1-c$}, a bit complementary to Alice's, with probability
\mbox{$\sin(2|y-r|)$}, otherwise he outputs \mbox{$b=c$} just like Alice.

The probability that $a=b$ is calculated as an integral over the various
possibilities for~$r$, as if it had been chosen after $x$ and $y$ are fixed.
For~simplicity, assume that \mbox{$x \le y$}.  The probability that
\mbox{$a=b$} is~1 if \mbox{$0 \le r \le x$} or \mbox{$y \le r \le 1$}
since, in that case, $r$~does not lie between $x$ and~$y$.
Otherwise, if \mbox{$x<r<y$}, the probability that \mbox{$a=b$} is
\mbox{$1-\sin (2(y-r))$}.  Therefore, the global probability that \mbox{$a=b$}
is
\[ \int_{r=0}^{x} 1\,\dee r
+ \int_{r=x}^{y} [ 1-\sin(2(y-r)) ] \,\dee r
+ \int_{r=y}^{1} 1\,\dee r \]
\[ = \half + \half \cos(2(y-x))  =  \cos^2(x-y) \, ,\]
as required.
It~is tempting to ``improve'' on this approach and make it work for all angles
between $0$ and $\pi$, simply with an appropriate change in the probability
function \mbox{$\sin (2|y-r|)$} that determines whether or not Bob will
output the same bit as Alice when $r$ lies between $x$ and~$y$.
Unfortunately, any such attempt will result in ``probabilities''
that are either negative or greater than~1\,!

It is shown in~\cite{BCT} how to simulate an arbitrary von Neumann
measurement with only \emph{eight} classical bits of communication in the
worst case, but it is left as an open question to determine whether or not the
effect of an arbitrary positive-operator-valued measurement (\textsc{povm}) can
be simulated with a bounded amount of classical communication in the worst case.

A different approach to the classical simulation of entanglement was taken
independently by Michael Steiner~\cite{steiner}, who showed that the EPR
relation can be simulated exactly with significantly fewer \emph{expected} bits
of classical communication, provided we accept that there be no
upper limit on the required amount
of communication in the case of bad luck. Steiner's technique was subsequently
refined by Nicolas Cerf, Nicolas Gisin and Serge Massar~\cite{CGM},
who showed that as few as 1.19 expected bits of classical communication suffice
to simulate exactly an arbitrary von Neumann measurement.
Even an arbitrary \textsc{povm} can be simulated by this technique,
at the expected cost of 6.38 bits of classical communication.

Building on~\mbox{\cite{steiner,CGM}}, Serge Massar, Dave Bacon, Nicolas Cerf
and Richard Cleve~\cite{MBCC} discovered that the exact classical simulation
of quantum entanglement can be achieved without any need for shared random
variables, provided we are satisfied with an \emph{expected} bounded amount of
classical communication.  In~particular, they show how to simulate the
effect of an arbitrary \textsc{povm} on one bit of entanglement with less than
20 bits of expected classical communication.  Conversely, they also show that
the exact simulation of quantum entanglement with a worst-case bounded amount of
classical communication (as in the scenario of~\cite{BCT} described earlier in
this section) is \emph{not} possible without an infinite amount of shared
randomness.

Finally, the question asked at the beginning of this section is almost
resolved in~\cite{BCT}.  It~is still unknown if there exists a constant~$c$
such that $c2^k$ bits of classical communication are sufficient to simulate
exactly the effect of $k$ bits of entanglement for all values of~$k$.
However, it is shown in~\cite{BCT} that as few as \mbox{$(3k+6)2^k$}
expected bits of classical communication suffice to simulate the
outcome of any \textsc{povm} that Alice and Bob could perform on
their respective shares of $k$ bits of entanglement.
This simulation protocol does not require Alice and Bob to share
random variables in the initialization phase.  No~similar results are known
for worst-case bounded communication even if we allow the sharing of
continuous random variables.

\section{Conclusions and Open Problems}\label{concl}

We have seen a variety of scenarios according to which quantum mechanics
allows for a significant improvement in the efficiency of communication,
compared to what would be possible in a classical world.  This is surprising
because the transmission of $n$ quantum bits cannot serve to communicate more
than $n$ classical bits of information, and because quantum entanglement on
its own cannot be used to communicate at all.  Perhaps the most interesting
aspect of quantum communication complexity is that the advantage provided
by quantum mechanics has been established rigourously.  This is in sharp
contrast with the field of quantum \emph{computing}, in which it is merely
\emph{believed} that quantum mechanics allows for an exponential speedup in
some computational tasks, such as the factorization of large
numbers~\cite{shor}. Indeed, it has not yet been ruled out that there might
exist an efficient factorization algorithm for the classical computer.

Several interesting questions are still open.  The exponential advantage
of quantum communication over classical communication
has been established only in the case of promise problems, in both
the error-free~\cite{BCW} and bounded-error~\cite{raz} scenarios.
Could there be a (total) function \mbox{$f:X \times Y \rightarrow Z$},
where \mbox{$X=Y=\{0,1\}^n$}, for which the distributed computation of
$f$ would be exponentially more efficient with quantum communication
compared to classical communication?

We have seen at the end of Section~\ref{alaCB} that the amount of
classical communication required for the accomplishment of a
distributed task in the presence of unlimited entanglement
cannot be more than twice the amount of quantum communication that
would suffice for the same task, because quantum teleportation
can be used to transmit quantum bits through a classical channel.
How about the other direction?
Could there be a task that can be accomplished with a small amount
of classical communication in the presence of unlimited entanglement,
but that would require a much larger amount of quantum communication
if prior entanglement were not available?

We have seen in Section~\ref{spooky} that the Deutsch--Jozsa relation
can be established classically without communication when \mbox{$n=2$}
or \mbox{$n=4$}, but not when $n$ is
arbitrarily large.  But how large must ``large''~be?
In~particular, can it be established for~\mbox{$n=8$}?
It~is~\mbox{interesting} to note that the Deutsch--Jozsa relation becomes
easier and easier to \emph{fake} when $n$ becomes larger.  Indeed, if Alice
and Bob share $k$ random variables
\mbox{$t_1$, $t_2$, \ldots, $t_k \in \{0,1\}^n$} in the initialization phase,
and if they output \mbox{$a_i = x \cdot t_i$}, and
\mbox{$b_i = y \cdot t_i$}, respectively, their probability of being caught
with \mbox{$a=b$} if in fact \mbox{$\Delta(x,y)=n/2$} goes down as~$2^{-k}$.
A~nice open question is to determine the task on $n$ input and $k$
output bits that can be handled with certainty given sufficient entanglement
and no communication,
but for which the probability of success would be as small as possible
in a classical world.

Finally, several open questions are given in Section~\ref{simul} concerning
the classical simulation of quantum entanglement.
Is~it possible to achieve the EPR task with fewer than four bits
of classical communication in the worst case? 
Is~it possible to simulate an arbitrary \textsc{povm} with a \emph{worst-case}
bounded amount of classical communication?
How much classical communication is sufficient in the worst case to simulate
the effect of $k$ bits of entanglement?
In~the expected case?
We~have seen how classical communication can be used to simulate
entanglement for tasks that did not involve classical communication
in the quantum setting.  How about the classical simulation of
tasks that use not only quantum entanglement but also classical
(or perhaps quantum) communication?

\newpage

\section*{Acknowledgements}

I am grateful to Richard Cleve for introducing me to the fascinating
field of quantum communication complexity and to Alain Tapp for sharing with
me some of his best thoughts on the topic.  In~addition to discussions with
Richard and Alain, this paper has benefited from insightful comments from
Harry Buhrman and Ronald de Wolf.

\end{document}